\documentclass[reprint,amsmath,amssymb,aps,twocolumn]{revtex4}

\usepackage{graphicx}
\usepackage{float}

\usepackage{bm}
\usepackage{color,soul}

\usepackage[normalem]{ulem}
\usepackage{xpatch}

\newcommand{\be}{\begin{eqnarray}}
\newcommand{\ee}{\end{eqnarray}}
\newcommand{\e}[1]{\textrm{e}^{#1}}

\begin{document}

\title{Master equation for non-Markovian quantum Brownian motion:\\The emergence of {\em lateral} coherences}

\author{Sapphire Lally\textsuperscript{1}}
\thanks{Equal contribution to this work}
\author{Nicholas Werren\textsuperscript{1,2}}
\thanks{Equal contribution to this work}
\author{Jim Al-Khalili\textsuperscript{1}}
\author{Andrea Rocco\textsuperscript{1,2}}
\email{a.rocco@surrey.ac.uk}

\affiliation{
 \textsuperscript{1}Department of Physics and \textsuperscript{2}Department of Microbial Sciences,  \\University of Surrey, GU2 7XH, Guildford, UK
}

\date{\today}

\begin{abstract}
  Understanding the behaviour of a quantum system coupled to its environment is of fundamental interest in the general field of quantum technologies. It also has important repercussions on foundational problems in physics, such as the process of decoherence and the so-called quantum measurement problem. There have been many approaches to explore Markovian and non-Markovian dynamics within the framework of open quantum systems, but the richness of the ensuing dynamics is still not fully understood. In this paper we develop a non-Markovian extension of the standard Caldeira-Leggett model, based on expanding the dynamics of the reduced system at high temperature in inverse powers of the high-frequency cut-off of the Ohmic spectral density of the environment, and derive a non-Markovian master equation for the reduced density matrix for the case of a general potential. We also obtain a fully analytical solution in the free particle case. While the short-time behaviour of this solution does not diverge substantially from the Markovian behaviour, at intermediate times we find a resurgence of coherence, which we name {\em lateral coherence}. We identify this with a corresponding transient negative entropy production rate, which is understood to be characteristic of non-Markovian dynamics. We also analyze the positivity of the reduced density matrix, and derive the corresponding Fokker-Planck equation in the classical limit.
\end{abstract}

\pacs{Valid PACS appear here}

\maketitle

\section{Introduction}

The study of quantum systems in interaction with their environment -- so-called open quantum systems -- touches upon a number of unsolved problems in fundamental physics. Decoherence due to the interaction of a quantum system with its environment must be understood or controlled in all applications where non-classical dynamics need to be harnessed \cite{Joos96, Zurek03, Ladd10}. Much effort is thus currently devoted to either decoupling the system from its environment, or to engineering environmental features that protect quantum behaviour \cite{Fagaly10, Baumgart16}. 

Solving the dynamics of open quantum systems is, in general, a challenging task. A Markovian system is one in which the state of the system at time $t_n$ is uniquely determined by its state at time $t_{n-1}$, and not by earlier times. In the Markovian case, many different techniques have been developed to analyse the dynamics (see for instance \cite{deVega17} for a review). In contrast, a non-Markovian system retains memory of the state at earlier times, and the analysis is usually more involved.

Non-Markovian behaviour emerges naturally in both classical and quantum systems when the dynamics of environmental degrees of freedom are projected out. Both the adoption of general projection techniques \cite{Mori65,Nakajima58,Zwanzig60} and the explicit calculation of specific environmental models \cite{Feynman63} lead to reduced dynamics for the system of interest characterized by memory kernels which account in principle for non-Markovian behaviour. However, if {\em ad-hoc} assumptions about the relaxation timescales of such memory kernels are made, a Markov approximation can be performed. This leads to reduced quantum dynamics which are again Markovian, and characterized by very fast (approximately exponential) decoherence of the coherent components \cite{Zurek03}, accounting thereby for the environment-induced measurement process \cite{Presilla96,Presilla97}, and the quantum-classical transition \cite{Kovacs17}. 

On the other hand, since the pioneering work of Hu, Paz and Zhang \cite{Hu92},  investigation of non-Markovian dynamics continues to gain increasing attention in an attempt to obtain a deeper insight into quantum behaviours and in the study of the transition to classical dynamics \cite{Breuer08,Fleming11,Chin12,Laine14}. One important implication of non-Markovianity has emerged recently in the context of the resulting quantum thermodynamics and on its implications for novel entropy dynamics. Recent work has identified unexpected behaviour related to quantum coherences flowing back from the thermal bath into the system of interest \cite{Cai18, Kallush19}, a phenomenon which has been associated with local negative entropy variations \cite{Strasberg19, Rivas20, Popovic18, Bhattacharya17, Marcantoni17}, and even heat flow reversal \cite{Latune19, Latune20, Micadei19}. This type of surprising behaviour was recently given a thermodynamic and information theoretical justification in \cite{Ahmadi21}: the information backflow that characterises non-Markovianity \cite{Mazzola10, Bylicka16, Lu10, Chruscinski11} is used by the system to ``decode" lost information about its internal degrees of freedom and thereby, in principle, allow the system to perform more thermodynamic work than its Markovian counterpart.

In this paper we address non-Markovian dynamics by building upon the path integral approach of Caldeira and Leggett \cite{Caldeira83}. We model the environment as a bath of oscillators at high temperature with a high frequency cut-off, $\Omega$, and develop a perturbative expansion in powers of $\Omega^{-1}$. This method is similar to that applied in \cite{Fraga14}, in which the same expansion is applied to derive the classical Langevin equation. We obtain a quantum master equation with non-Markovian corrections perturbatively added to the standard Lindblad Markovian structure \cite{Lindblad76}. It is noticeable that the perturbative method introduced here converts the intrinsic time non-locality of the non-Markovian regime \cite{Hu92,Fleming11b} into a local structure defined in terms of higher order spatial derivatives, thereby allowing an easier interpretation of the reduced dynamics. Our method is also valid for any general scalar potential felt by the system of interest. 

We also present an exact analytical solution of the master equation, based on the method developed in \cite{Roy99}. Our solution shows the presence of an increase of coherence of the system of interest at intermediate times. After a fast decoherence process, not too dissimilar from that occurring in the standard Markovian system, our solution predicts a resurgence of quantum coherences, characterised by non-zero density matrix elements away from the diagonal. The phenomenon is weakly observed on the decay of the initial coherent modes of the reduced density matrix, and quite pronounced when it emerges from the population structure. For this reason we name these \textit{lateral coherences}. We demonstrate that there is a corresponding decrease in the von Neumann entropy of the system of interest, associated with a non-Markovian backflow of information into the system, in agreement with \cite{Marcantoni17, Bhattacharya17}.

We also study the positivity of the non-Markovian master equation here derived. Positivity violation, with the consequent violation of the uncertainty principle, is a longstanding issue in the general analysis of open quantum systems, which manifests itself in the density matrix spectrum characterized by at least one negative eigenvalues. While all open quantum systems falling in the Lindblad class can be shown to preserve positivity \cite{Lindblad76,Gorini76}, other models, including the Caldeira and Leggett model, show violations of it, which have been analysed in detail since the pioneering works of Dekker \cite{Dekker77} and Di\'osi \cite{Diosi93}. The master equation that we obtain is also expected to show some positivity violations \cite{Kohler97} on general grounds, but we show that these do not occur in the parameter regime analyzed here. We adopt the approach of \cite{Homa19}, which has highlighted that the purity of the system and the Robertson-Schr\"odinger uncertainty relation provide a necessary condition for positivity of the density matrix to be preserved. Numerical evaluation of these quantities for our non-Markovian system indeed show that these stay positive, giving us confidence in the physical reliability of our derivation and conclusions.

Finally, we derive the evolution equation of the Wigner function, and the corresponding Fokker Planck Equation in the classical limit. We find that there is a new term, corresponding to non-Markovian dynamics in the classical regime. We also find that our non-Markovian description bears similarity to known results obtained in a different contexts \cite{Unruh89, Hu92, Kohler97,Dillenschneider09}.

In Section 2 we review the Caldeira and Leggett model. In Section 3 we present our novel perturbative expansion in terms of inverse powers of $\Omega$, in Section 4 we derive the analytical solution of the system, and in Section 5 we illustrate the obtained reduced dynamics. We also show the implications of the new non-Markovian dynamics on the decoherence process and emergence of novel lateral coherence at medium timescales. In Section 6 we discuss the positivity of the reduced density matrix, and in Section 7 we obtain the classical Fokker-Planck equation for the reduced system. Finally we draw conclusions in Section 8.

\section{The Caldeira-Leggett master equation}

In the Caldeira and Leggett Model \cite{Caldeira83} the dynamics of the system are described by the total Lagrangian
\be
\mathcal{L} = \mathcal{L}_{A} + \mathcal{L}_{B} + \mathcal{L}_{I}\ ,
\ee
where 
\be
&&\mathcal{L}_{A} = \frac{1}{2} M \dot x^{2} - V(x)\ ,\\
&&\mathcal{L}_{B} = \sum_{k=1}^N \left(\frac{1}{2} m \dot R_{k}^{2} - \frac{1}{2} m \omega_{k}^{2} R_{k}^{2} \right)\ ,\\
&&\mathcal{L}_{I} = - x \sum _{k=1}^N C_k R_k\ ,
\ee
are the Lagrangians describing the system of interest, the bath oscillators, and their interaction, respectively. 

By using the path integral formalism adopted in \cite{Feynman63,Caldeira83}, the degrees of freedom of the bath can be integrated out, and the time evolution of the reduced density matrix of the system written in terms of the superpropagator
\be
\!\!\!J(x_{f},y_f,t;x_{i}, y_{i},0)\! =\!\!\! \int\!\! \!\!\int\!\!\mathcal{D}x \mathcal{D}y e^{i(S_A[x]-S_A[y])} \! {\cal F} \left[ x, y \right], \label{prop1}
\ee
where the above denote path integrals over all paths $x(t)$ ($y(t)$) 
from initial point $x_i=x(0)\left(y_i=y(0)\right)$ to final point $x_f=x(t)\left(y_f=y(t)\right)$, $S_A$ is the classical action of the system of interest, and ${\cal{F}} \left[ x, y \right]$ is the influence functional, which describes the action of the environment on the system of interest. The explicit structure of the influence functional can be computed as in \cite{Caldeira83},
\be 
&&{\cal F} \left[ x, y \right] = \exp \left[ - \frac{i}{\hbar} \int^{t}_{0} d t^{\prime} f_{-}(t^{\prime}) I_I(t^{\prime}) \right. \nonumber \\
&&\hspace{3cm} \left. - \frac{1}{\hbar} \int^{t}_{0} dt^{\prime} f_{-}(t^{\prime}) I_R(t^{\prime}) \right]\!,
\ee
where $I_I(t^{\prime})$ and $I_R(t^{\prime}) $ are convolution integrals, given by 
\be
&&I_I(t^{\prime}) = \int^{t^{\prime}}_{0} \alpha_{I}(t^{\prime} - t^{\prime\prime}) f_{+}(t^{\prime\prime}) dt^{\prime\prime}, \label{conv1}\\ 
&&I_R(t^{\prime}) = \int^{t^{\prime}}_{0} \alpha_{R}(t^{\prime} - t^{\prime\prime}) f_{-}(t^{\prime\prime}) dt^{\prime\prime}, \label{conv2}
\ee
and $f_{\pm}(t) = x(t) \pm y(t)$. The integrals (\ref{conv1}) and (\ref{conv2}) are specified in terms of the memory kernels 
$\alpha_{I}$ and $\alpha_{R}$, which in the high temperature limit $k_B T \gg \hbar \omega_k$ and under the assumption of an Ohmic environment, read \cite{Caldeira83}:
\be 
&&\alpha_I(t) = \frac{2M \gamma}{\pi} \frac{d}{dt}
\left(\frac{\sin \Omega t}{t}\right), \label{kernelIm} \\
&&\alpha_R(t)= \frac{4 M \gamma k_B T}{\pi \hbar}
\frac{\sin \Omega t}{t}. \label{kernelRe}
\ee
Here, $\gamma$ is the relaxation constant of the system of interest, and $\Omega$ is the cut-off frequency of the spectral density of the bath oscillators. This emerging non-Markovian structure, contained here by the convolution integrals, is a common occurrence in the reduction of both classical and quantum systems \cite{Hanggi78, Chruscinski10}, stemming from projecting out the environmental degrees of freedom. 

In the Markov limit $(\Omega \rightarrow \infty)$, the memory kernels (\ref{kernelIm}) and (\ref{kernelRe}) become delta functions and a Markovian representation for the reduced dynamics of the system of interest is recovered. In this case, the reduced density matrix $\tilde{\rho}$ obeys the well-known master equation \cite{Zurek03}:
\be
\frac{\partial \tilde{\rho}}{\partial t} = \hat{L}_{\text{M}} \tilde{\rho} \ ,\label{zurek}
\ee
where $\hat{L}_{\text{M}}$ is the Liouvillian operator in the Markov limit defined as
\be
&&\hspace*{-1.cm} \hat{L}_{\text{M}} \tilde{\rho} =-\frac{i}{\hbar} \left[\hat{H}_R,\tilde{\rho} \right] \nonumber \\
&&\hspace*{-0.5cm} -\gamma\left[ (x-y) \left(\frac{\partial }{\partial x}  -  \frac{\partial }{\partial y} \right) +  \frac{2Mk_BT}{\hbar^2}(x-y)^2\right] \tilde{\rho}.\label{Liouvillian}
\ee
Here, $\hat{H}_R$ is the renormalized Hamiltonian of the system. The second and third terms in (\ref{Liouvillian}) represent relaxation and decoherence dynamics, respectively.

\section{A perturbative non-Markovian expansion}

We now develop a time-scale expansion in inverse powers of $\Omega$ and identify perturbatively the contribution of terms in the neighbourhood of $\Omega=\infty$. In order to obtain an expansion of the integrals (\ref{conv1}) and (\ref{conv2}) in powers of $1/\Omega$, we make use of Laplace Transforms, which, for $\Omega \gg |s|$, allow us to write
\be
\hspace{-0.5cm}\mathcal{L} \left \{   \frac{\sin (\Omega t)}{t} \right\}(s) &=& \sum^{\infty}_{n=0} \left( \frac{(-1)^{n} \left( \frac{\Omega}{s} \right)^{2n+1}}{(2n+1)} \right) \nonumber \\
&=& \arctan\left(\frac{\Omega}{s}\right) = \frac{\pi}{2} - \arctan\left(\frac{s}{\Omega}\right) \nonumber \\
&=& \frac{\pi}{2} - \sum^{\infty}_{n=0} \left( \frac{(-1)^{n} \left( \frac{s}{\Omega} \right)^{2n+1}}{(2n+1)} \right)\!.
\ee
By using the Convolution Theorem for Laplace Transforms, and after inverse transforming back to the time domain, we obtain the corresponding expression for the superpropagator which, to first order in $1/\Omega$, reads: 
\be 
&&\!\!\!\!\!\!\!\!\! J(x_{f}, y_{f},t;x_{i}, y_{i},0) =  \int \!\!\!\! \int \mathcal{D}x \mathcal{D}y\, C_\Omega(t) \nonumber \\ && \hspace{2.5cm}\; e^{\frac{i}{\hbar}\left(S_A[x] - S_A[y]\right)} \; {\cal F}^{(\Omega)}[x,y] , \label{Prop2} 
\ee
where $C_{\Omega}(t)$ is a prefactor collecting all boundary terms,
\be
&&\hspace{-0.5cm}C_\Omega(t) = \exp \left[\int^{t}_{0} d\tau \bigg( -\frac{i}{\hbar}\left(\frac{2M \gamma }{\pi}\right) \left(\frac{\pi}{2} \delta(\tau)f_{-}(\tau)f_{+}(\tau)   \right. \right.\nonumber \\
&&\hspace{-0.5cm} - \left. \left. \frac{1}{\Omega} {\delta}^{\prime}(\tau)f_{-}(\tau)f_{+}(0) - \frac{1}{\Omega}  \delta^{\prime}(\tau)f_{-}(\tau) f_{+}^{\prime}(0) + {\cal O} (\Omega^{-2}) \right) \bigg) \right]   \nonumber \\  
&&\hspace{-0.5cm} \times \exp \left[\int^{t}_{0} d\tau \bigg( - \frac{1}{\hbar} \left(\frac{4 M \gamma k T}{\hbar \pi}\right) \right. \nonumber \\
&&\hspace{-0.5cm} \times \left.\left(\frac{1}{\Omega} \delta(\tau) f_{-}(\tau)f_{-}(0) + {\cal O} (\Omega^{-2})\right)\bigg)\right] 
\ee
and the influence functional is now given by 
\be
\label{eq:inffuncNM}
&& {\cal F}^{(\Omega)}[x,y] =\nonumber \\
&&\exp\left[ - \frac{i}{\hbar} \left(\frac{2M \gamma }{\pi}\right) \int^{t}_{0} dt^{\prime} \left( - \frac{\pi}{2}  f_{-}(t^{\prime}) \frac{df_{+}(t^{\prime})}{dt^{\prime}} \right. \right.  \nonumber \\
  && \hspace{0cm}- \Omega f_{-}(t^{\prime})  f_{+}(t^{\prime}) + \left. \left. \frac{1}{\Omega} f_{-}(t^{\prime}) f_{+}^{\prime \prime}(t^{\prime}) + {\cal O} (\Omega^{-2})  \right) \right. \nonumber \\
&& \hspace{0cm}\left. - \frac{1}{\hbar} \left( \frac{4 M \gamma k_B T}{\hbar \pi}\right) \int^{t}_{0} dt^{\prime} \left( -\frac{\pi}{2}f_{-}^2(t^{\prime}) \right. \right. \nonumber \\
&& \hspace{0cm} + \left. \left. \frac{1}{\Omega}  f_{-}(t^{\prime}) f_{-}^{\prime}(t^{\prime}) + {\cal O} (\Omega^{-2}) \right) \right],
\ee
where $f_{\pm}(t) = x(t) \pm y(t)$ as before. In order to compute the time evolution of the reduced density matrix $\tilde{\rho}$, we now consider:
\be 
&&\hspace{-0.5cm}\tilde{\rho}(x_{f}, y_{f},t) =\nonumber \\
&&\hspace{0.3cm}\int\!\!\!\!\int \!\! dx_{i} dy_{i}\, J(x_{f}, y_{f},t;x_{i}, y_{i},0)\, \tilde{\rho}\left(x_{i}, y_{i},0 \right). \label{QMENMProp}
\ee
We follow the general principles of the standard approach, based on considering the propagation over a small interval of time $\varepsilon$. However, here we need to consider the propagation over two intervals to accommodate the effects of the second derivative in the superpropagator. Therefore we write 
\be 
&&\tilde{\rho}(x, y, t + 2 \epsilon) = \\ \nonumber && \hspace{0.3cm}\int\!\!\!\!\int \!\! dx_0 dx_1 dy_0 dy_1\, J(x, y, t + 2\epsilon;x_0, y_0,0)\, \tilde{\rho}\left(x_0, y_0,t \right),
\ee
where $x_1 = x(t + \epsilon)$, $y_1 = y(t + \epsilon)$ are new ``intermediate" variables that must also be integrated out. The superpropagator separates into two contributions: The first, $J_{\textrm{QM}}$, is a ``quasi-Markovian" part, which obeys the Chapman-Kolmogorov equation \cite{Breuer16} - the integrand is separable in $x_1$, $x_0$ and $y_1$, $y_0$ - and can therefore be integrated over each interval independently. The second contribution, $J_{\textrm{NM}}$, is a truly non-Markovian part, which cannot be separated over the intervals:
\be
&&\hspace{-0.3cm} \tilde{\rho}(x, y, t + 2 \epsilon) = \\ \nonumber 
&&\hspace{-0.3cm}\int\!\!\!\!\int \!\! dx_1 dy_1 \, \Bigg( J_{\textrm{QM}}(x, y, t + 2\epsilon;x_1, y_1,t+\epsilon)\, \\ \nonumber
&& \hspace{-0.3cm}\times \int\!\!\!\!\int \!\! dx_0 dy_0 \, J_{\textrm{QM}}(x_1, y_1, t + \epsilon;x_0, y_0, t)\, \tilde{\rho}\left(x_1, y_1,t+\epsilon \right)\Bigg) \\ \nonumber
&&\hspace{-0.3cm} + \int\!\!\!\!\int \!\! dx_0 dx_1 dy_0 dy_1\, J_{\textrm{NM}}(x, y, t + 2\epsilon;x_0, y_0,0)\, \tilde{\rho}\left(x_0, y_0,0 \right)
\ee
The quasi-Markovian and non-Markovian integrals are constructed by making the finite-difference substitutions 
\be 
&& \dot{x} \approx \frac{x_1 - x_0}{\varepsilon} = \frac{\beta_{x_0}}{\epsilon} \\ \nonumber 
&& \ddot{x} \approx \frac{(x_2 - x_1) - (x_1 - x_0)}{\varepsilon^2}= \frac{\beta_{x_1} - \beta_{x_0}}{\varepsilon^2},
\ee
and the corresponding expressions for $\dot{y}, \ddot{y}$, into Eq. (\ref{eq:inffuncNM}). Then, it is a straightforward but lenghty procedure to integrate over the $\beta$-variables, truncating terms that are $O(\varepsilon^2)$ or $O(1/{\Omega^2})$. The resulting expression can be organised in powers of $\varepsilon$, and by collecting terms linear in $\varepsilon$, we obtain a new master equation for the density matrix, to leading order in our non-Markovian expansion in $1/\Omega$:
\be
\hspace{-0.6cm} \frac{\partial \tilde{\rho}}{\partial t} &=& \left[ \hat{L}_{\text{M}} 
+ \frac{\gamma}{\pi \Omega} \left(\hat{L}_{\text{R}}\! +\!\hat{L}_{\text{V}}\! +\!\hat{L}_{\text{O}}\! \right) \right] \tilde{\rho} ,\label{nonMME}
\ee
where $\hat{L}_{\text{M}}$ is the Markovian Liouvillian defined in Eq. (\ref{Liouvillian}) and the new, non-Markovian, terms are
\be
&&\hspace{-0.3cm} \hat{L}_{\text{R}} = - 2  \gamma (x-y) \left(\frac{\partial }{\partial x}  -  \frac{\partial}{\partial y} \right) \, \label{nonMME1} \\ 
&&\hspace{-0.3cm} \hat{L}_{\text{V}} =  -\frac{i}{2 \hbar} (x-y) \left(\frac{dV}{dx} + \frac{dV}{dy} \right)  \, \label{nonMME2} \\
&&\hspace{-0.3cm} \hat{L}_{\text{O}} =  -\frac{4 i  k_BT}{\hbar} (x-y) \left(\frac{\partial }{\partial x} + \frac{\partial}{\partial y} \right) \ . \label{nonMME3}
\ee

In contrast to the HPZ master equation \cite{Hu92}, equation (\ref{nonMME}) trades off explicit time dependencies of coefficients in favour of dynamical operators acting on the reduced density matrix, and therefore offers a clearer physical interpretation of the ensuing time evolution of the system. The first Liouvillian operator, Eq. (\ref{nonMME1}), is of the same form as the standard relaxation term in Zurek equation, and therefore it only provides with a cut-off dependent renormalization of the related coefficient. The second Liouvillian, Eq. (\ref{nonMME2}), is a potential dependent term, whose effect needs to be assessed case by case. The most interesting Liouvillian is the last one, Eq. (\ref{nonMME3}). As we shall see it corresponds to a term in the evolution equation for the Wigner function already identified in a different context \cite{Dillenschneider09}. It here emerges as a truly non-Markovian contribution to the evolution of the density matrix, and its effect is far from being trivial, as we show in the next section.   

We note that Eq. (\ref{nonMME}) reduces to the Markovian master equation (\ref{zurek}) for $\gamma/\Omega \rightarrow 0$. Because of the structure of Eq. (\ref{nonMME}), it is natural to define the dimensionless parameter $R_{\Omega} = \gamma/(\pi \Omega)$ as a measure of non-Markovianity of the system. If we identify $\tau_B \sim 1/\Omega$ as the timescale of the environment (corresponding to the period of the fastest, most dominant, bath oscillator) and $\tau_A = 1/\gamma$ as the relaxation timescale of the system of interest, the Markovian condition $\Omega \gg \gamma$ corresponds to $\tau_B \ll \tau_A$, and hence justifies the reduction of the dynamics obtained by integrating over the fast degrees of freedom of the environment. The case $\Omega \sim \gamma$ represents a situation in which system and environment become indistinguishable. 
The analysis of \cite{Einsiedler20} thoroughly investigates the parameter space created by $\{T, \gamma, \Omega\}$ in the context of non-Markovian quantum Brownian motion, by using the trace distance measure of non-Markovianity \cite{Breuer09} to identify optimal combinations of parameter values. Although we are limited to $\Omega \gg \gamma$ by our perturbative expansion, this work suggests some natural choices of values for the parameters of interest.

\section{Exact analytical solution}

We approach the derivation of the analytical solution by following the approach laid out in \cite{Roy99}. For the free particle, the master equation can be written as 
\be 
&&\hspace{-0.3cm}\frac{\partial}{\partial t} \tilde{\rho}(x, y, t) =\Big[\frac{i\hbar}{2 M}\left ( \frac{\partial^2}{\partial x^2} - \frac{\partial^2}{\partial y^2}\right) \\ \nonumber
&&\!- \gamma(1 + 2R_{\Omega}) (x-y)\left ( \frac{\partial}{\partial x} - \frac{\partial}{\partial y} \right ) - \frac{2 M \gamma k_B T}{\hbar^2}(x-y)^2 \\ \nonumber
&&\hspace{1cm}- \frac{4i R_{\Omega} k_B T}{\hbar} (x-y) \left ( \frac{\partial}{\partial x} + \frac{\partial}{\partial y} \right ) \Big]\tilde{\rho}(x, y, t) 
\ee
This is cast in canonical form via the coordinate transformation 
\be
    u = \frac{x + y}{2}, ~~~~ v = x-y,
\ee
and the rotated coordinate $u$ is then Fourier transformed into its associated momentum variable $K$,
such that we can write the master equation as a first order differential equation in one spatial dimension
\be
&& \frac{\partial}{\partial t} \tilde{\rho}(K, v, t) +\Big (\left(2\gamma_R v+ \frac{\hbar K}{m}\right) \frac{\partial}{\partial v} \\ \nonumber
&& \hspace{1cm} + \frac{2m \gamma k_B T}{\hbar^2} v^2 - \frac{4 R_{\Omega} k_B T}{\hbar} K v  \Big) \tilde{\rho}(K, v, t) = 0.
\ee
For convenience, we have written $\gamma_R = \gamma ( 1 + 2R_{\Omega})$.
This equation can then be solved via the method of characteristics by introducing a parameterisation variable $s$ such that $v = v(s)$, $t = t(s)$, and $\tilde{\rho} = \tilde{\rho}(s)$ to give the characteristic equations 
\be
    &&\frac{\partial t}{\partial s}  = 1 \\ \nonumber
    &&\frac{\partial v}{\partial s} = 2\gamma_R v+ \frac{\hbar K}{m} \\ \nonumber 
    &&\frac{\textrm{d} \tilde{\rho}}{\textrm{d} s} = -\Big(\frac{2m \gamma k_B T}{\hbar^2} v(s)^2 - \frac{4 R_{\Omega} k_B T}{\hbar} K v(s)  \Big) \tilde{\rho}(s).
\ee
By solving the equations for $t(s)$ and $v(s)$, with initial conditions $t(0) = 0$ and $v(0) = v_0$, we can substitute their solutions into the equation for $\tilde{\rho}(s)$, which is now a function of $s$ only. After solving for $\tilde{\rho}(s)$, substituting $t=s$, and inserting the definition of 
\be
&&\hspace{-0.9cm}v_0 = -\frac{\hbar K}{2m \gamma_R} + \left(v(t) + \frac{\hbar K}{2m \gamma_R (1 + 2R_{\Omega})} \right)\textrm{e}^{-2 \gamma_R t},
\ee
we finally obtain 
\be 
\label{eq:fouriersol}
&& \rho(K, v, t) = \rho_0(K, v_0) \times  \\ \nonumber
&& \textrm{exp} \Big\{ - \frac{ k_B T}{2 m \gamma_R (1 + 2R_{\Omega})} \Big( K^2 (1 + 4R_{\Omega}+8R_{\Omega}^2 ) t \\ \nonumber
&&- \frac{m K}{\hbar}(1 + 4 R_{\Omega} + 8R_{\Omega}^2)[v - v_0]+ \frac{m^2 \gamma_R}{\hbar^2}[v^2 - v_0^2] \Big )\Big\}
\ee
for the Fourier space solution. This is already a useful solution in its present form. The $R_{\Omega}\rightarrow0$ limit of Eq. (\ref{eq:fouriersol}) is the final step presented in \cite{Roy99}. However, when the initial condition has a Gaussian form, we can exploit the convenient Fourier transform of a Gaussian function to recover the solution in coordinate space. This is done by collecting the coefficients of $K$ in both the initial condition and the dynamical part of the Fourier-space solution and writing it as a sum of Gaussians in $K$ that can be inverse Fourier transformed using the standard identity 
\be 
&&\mathcal{F}_K^{-1} [\e{a_2 K^2 + a_1 K + a_0}](u) \\ \nonumber 
&& \hspace{2cm}= \frac{\e{a_0}}{\sqrt{- 2 a_2}} \exp{-\frac{(a_1 - iu)^2}{4 a_2}}.
\ee
For example, in Zurek's classic example of a ``Schrodinger cat state" \cite{Zurek03} (i.e. a double Gaussian with peak displacement $\pm b$ and variance $\frac{1}{\sqrt{2 a}}$) the solution is 
\be
&&\hspace{-0.7cm}\tilde{\rho}(x, y, t) = \frac{\textrm{e}^{E_0} \times \sum_i \textrm{e}^{C_0^i-\frac{(E_1 + C_1^i - \frac{i(x+y)}{2})^2}{4 (E_2 + C_2^i)}}}{\left (1 + \textrm{e}^{-2a b^2} \right) \sqrt{-16 \pi (E_2 + C_2)}}, \label{solution}
\ee
with the index $i$ counting over $\textrm{c}^{\pm}$ (the ``coherences") and $\textrm{p}^{\pm}$ (the ``populations"). The solution is a linear combination of the contributions coming from the evolution of the initial populations, or on-diagonal states, and the contributions from the evolution of the initial coherences, or off-diagonal states. The coefficients are
\be
&&\hspace{-0.9cm}C_2 = -\frac{1}{8 a} - \frac{a \hbar^2 }{8 M^2 \gamma_R^2} \left(1 - \e{- 2 \gamma_R t} \right)^2 \\ \nonumber
&&\hspace{-0.9cm}C_1^{\textrm{p} \pm} = \pm i b + \frac{a \hbar (x-y) }{2M \gamma_R}\e{-2 \gamma_R t}\left (1 - \e{- 2 \gamma_R t} \right) \\\nonumber
&&\hspace{-0.9cm}C_0^{\textrm{p} \pm} = -\frac{1}{2}a (x-y)^2 \e{-4 \gamma_R t},\\\nonumber
&&\hspace{-0.9cm}C_1^{\textrm{c} \pm} = \frac{a \hbar \left ( 1 - \e{- 2 \gamma_R t}\right) \left((x-y) \e{- 2 \gamma_R t} \pm 2b \right)}{2 M \gamma_R} \\\nonumber
&&\hspace{-0.9cm}C_0^{\textrm{c} \pm} = -\frac{1}{2}a ((x-y) \e{- 2 \gamma_R t} \pm 2b)^2 \\\nonumber
&&\hspace{-0.9cm}E_2 = -\frac{k_B T}{2M \gamma_R ( 1+ 2R_{\Omega})} \Big ((1 + 4R_{\Omega} + 8R_{\Omega}^2) t \\\nonumber
&&+ \frac{\e{-2\gamma_R t} - 1}{2 \gamma_R}(1 + 4R_{\Omega} + 8R_{\Omega}^2) - \frac{(\e{-2\gamma_R t} - 1)^2}{4\gamma_R } \Big), \\\nonumber
&&\hspace{-0.9cm}E_1 =-\frac{k_B T}{2M \gamma_R ( 1+ 2R_{\Omega})}\frac{M (x-y)}{\hbar} \Big (\e{-2\gamma_R t}(1 - \e{-2\gamma_R t})\\\nonumber
&&- (1 - 4R_{\Omega} - 8R_{\Omega}^2)(1 - \e{-2\gamma_R t}) \Big), \\\nonumber
&&\hspace{-0.9cm}E_0 = -\frac{k_B T}{2M \gamma_R ( 1+ 2R_{\Omega})}  \frac{M^2 \gamma_R (x-y)^2}{\hbar^2}(1 - \e{-4\gamma_R t}).
\ee
From here, we can also deduce the steady-state solution, by taking the $t \rightarrow \infty$ limit of the coefficients. This gives 
\be 
\tilde{\rho}_{\infty}(x - y) = \frac{1}{\sqrt{2 \pi}} \e{- \frac{M k_B T}{2(1 + 2R_{\Omega}) \hbar^2} (x-y)^2}.
\ee
In the $R_{\Omega}\rightarrow0$ limit, this becomes the well-known Caldeira-Leggett steady state solution \cite{Roy99, Grossman09}. \\

\section{Analysis of non-Markovian dynamics}

To investigate the time evolution of the non-Markovian dynamics, we study numerically the free particle analytical solution (\ref{solution}), starting with a double Gaussian initial condition \cite{Zurek03}. For convenience, we choose the particle to be a proton and the Gaussians to be separated by 5 Angstroms. In order to ensure we are seeing true non-Markovian effects, we select parameters in the spirit of reference \cite{Einsiedler20}. We also assume the environment to be at a temperature of $T=320$K, and take a value for the relaxation constant of $\gamma = k_B T/20\hbar \sim 0.002$fs, if not otherwise stated. The range of $\Omega$ we investigate is in the region highlighted in \cite{Einsiedler20} as maximising the effects of non-Markovianity. Our analysis is presented separately in the short and long time regimes. 

\subsection{Short time behaviour}

\begin{figure}[t] 
  \includegraphics[height=3cm]{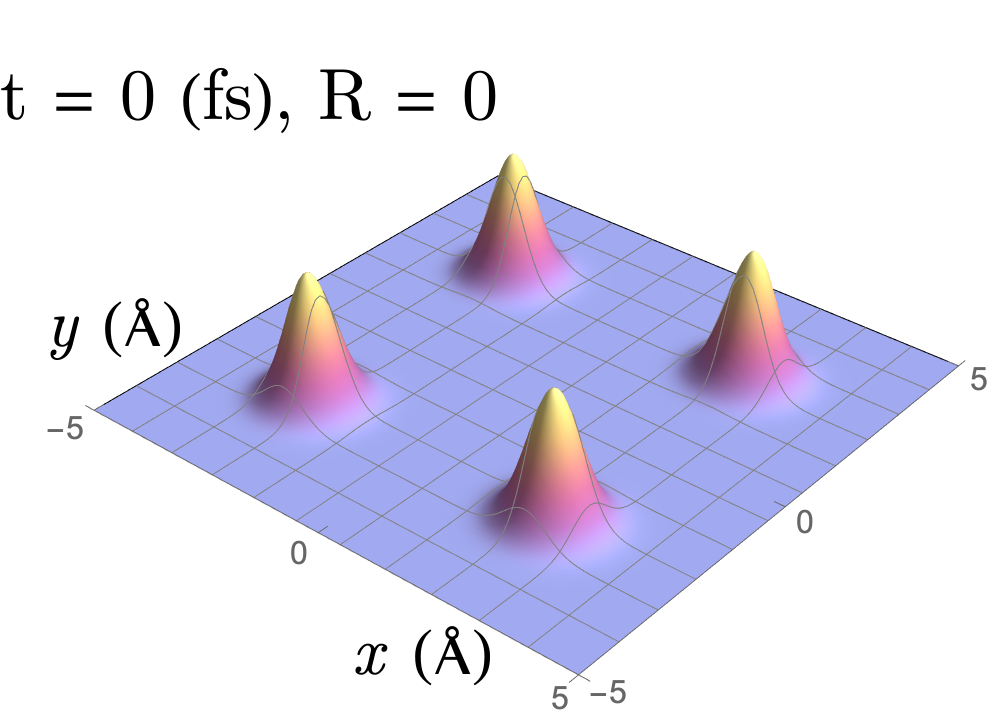} 
  \includegraphics[height=3cm]{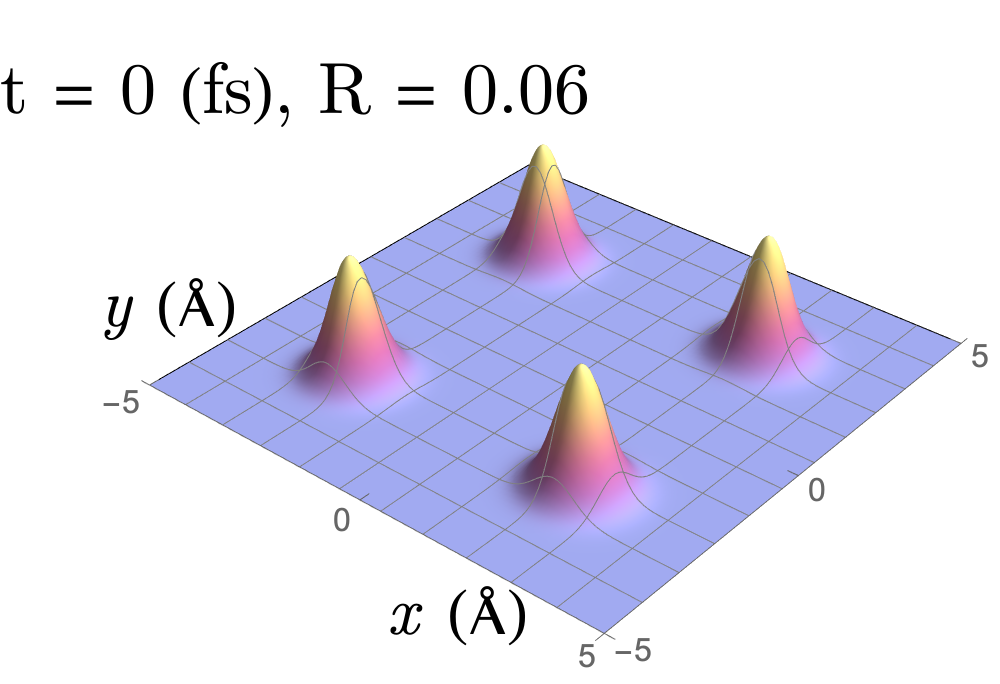}\\
  \includegraphics[height=3cm]{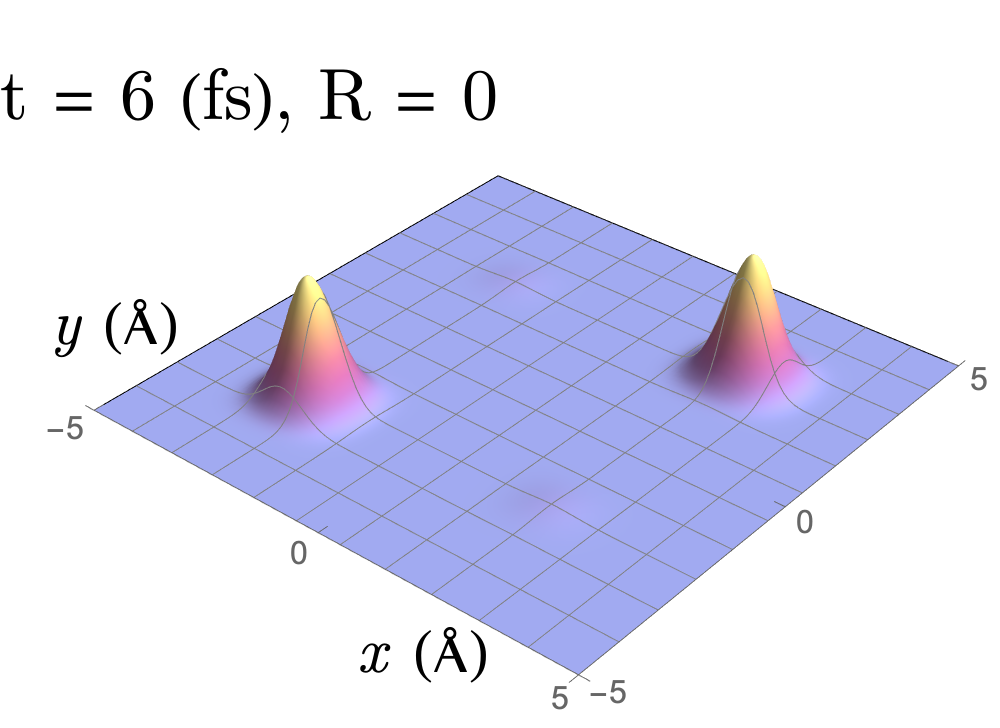} 
  \includegraphics[height=3cm]{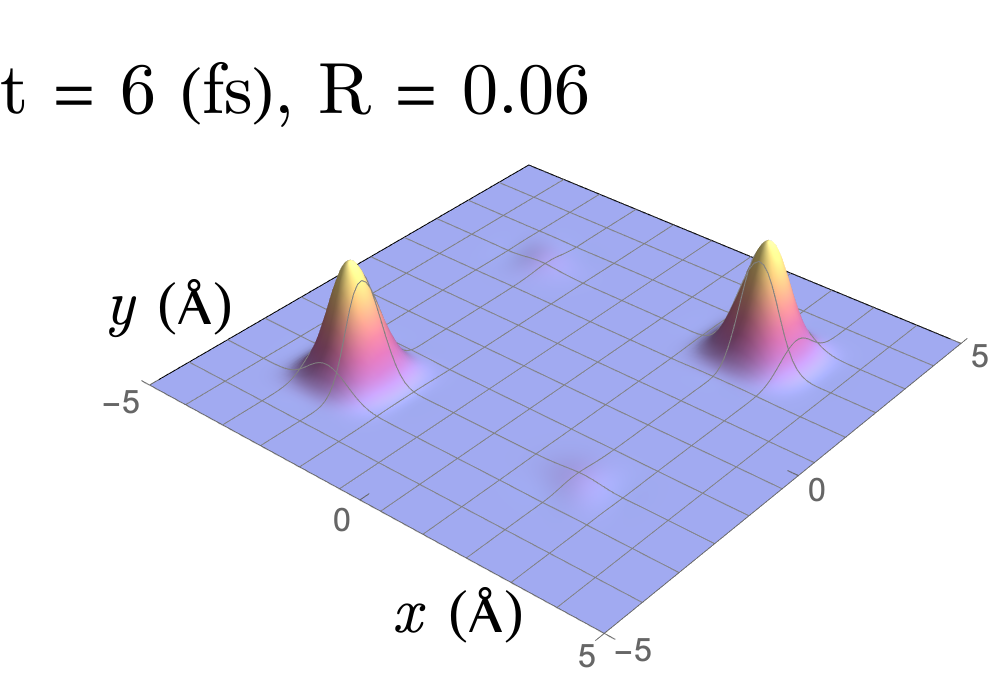}
  \caption{Short time evolution of the density matrix, via Eq. (\ref{solution}), in the Markovian case, $R_{\Omega}=0$ (left column), and the non-Markovian case, $R_{\Omega}=0.06$ (right column), for the free particle. Plots are over 6fs, to highlight the decoherence process. Parameters are $T=320 {\rm K}$, $\gamma = \frac{k T}{20 \hbar} {\rm fs}^{-1}$. Only a slight deviation between the Markovian and non-Markovian cases is apparent.} 
  \label{3DPlots_short}
  \end{figure}

We focus on analysing the non-Markovian behaviour on short time scales first. In Fig. \ref{3DPlots_short} we report the evolution of the density matrix up to 6 fs, so as to highlight the initial stages of the decoherence process. While the behaviour does not diverge substantially from the Markovian case, a slight delay in the decoherence process is visible.

To gain further insight into the decoherence process, we calculate the $l_1$-norm of coherence \cite{Baumgratz14}. For some choice of discrete basis, $\left\{\lvert i\rangle\right\}$, this is defined as $C_{l_1}=\sum_{i,j\neq i} \lvert \rho_{i,j}\lvert$. In the continuum limit of our configuration space basis, this becomes
\be 
C_{l_1}(t) = \int | \tilde{\rho}_{\rm off}(x,y,t) |\ dx\, dy\ . \label{coheq}
\ee
In a discrete system, it is straightforward to remove the $i=j$ diagonal elements from the sum. However, here the reduced density matrix is a continuous function of $x$ and $y$, and defining the diagonal elements is less straightforward. We take a density matrix without initial off-diagonal peaks, evolving under the Markovian Caldeira-Leggett / Zurek master equation, to be an incoherent ``baseline" which we can remove from our density matrix, such that $\tilde{\rho}_{\rm off} = \tilde{\rho} - \tilde{\rho}^{\rm CL}_{\rm decoh}$. 

\begin{figure}[t] 
  \includegraphics[height=5.5 cm, angle=0]{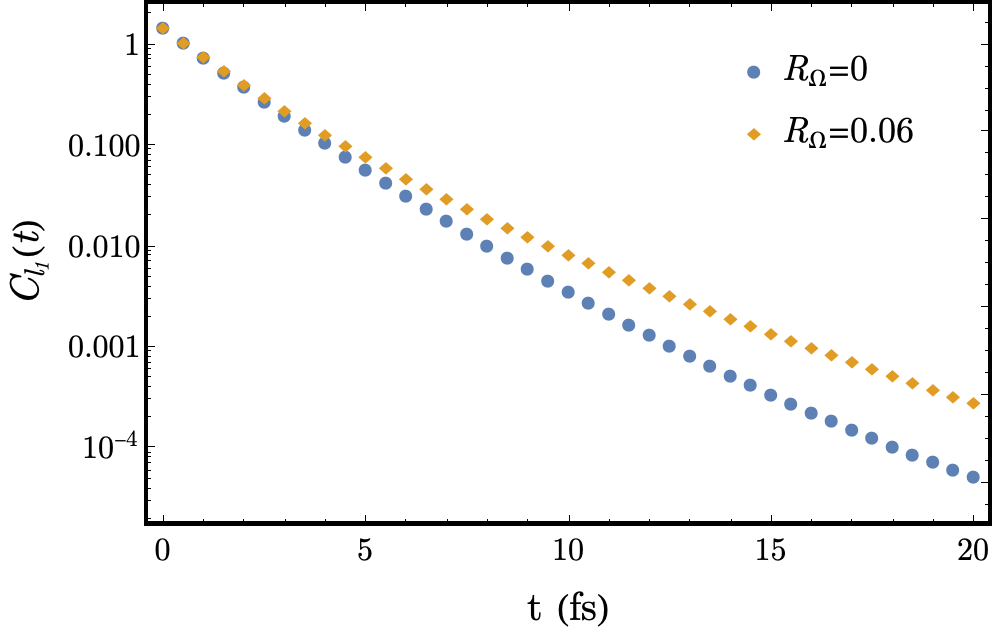}
  \caption{The $C_{l_1}$ norm of coherence, presented on a log-lin scale over $20 {\rm fs}$ in the Markovian $R_{\Omega}=0$ and non-Markovian $R_{\Omega}=0.06$ case. Parameters are $T=320 {\rm K}$, $\gamma = \frac{k T}{20 \hbar} {\rm fs}^{-1}$. The off-diagonals only have been isolated in this calculation to remove any effects from the diagonal. This calculation confirms that there is a slight but measurable deviation in the decoherence process.}\label{decbehaviourshort}
\end{figure}

In Fig. \ref{decbehaviourshort} we show that a slight delay in the decoherence process does indeed occur, which we ascribe to the presence of the Liouvillian $\hat{L}_{\text{O}}$, as the the decoherence term in Zurek equation is not modified by the finite cut-off frequency $\Omega$. As we will show in the next section, the action of $\hat{L}_{\text{O}}$ is to spread the peaks in an oscillatory manner. Although the effect is small, the off-diagonal peaks are spread along the line parallel to the diagonal, and this fractionally slows the decoherence process.

\subsection{Long time behaviour}

\begin{figure}[t] 
  \includegraphics[height=2.9cm]{r0-t0.png} 
  \includegraphics[height=2.9cm]{r006-t0.png}\\
  \includegraphics[height=2.9cm]{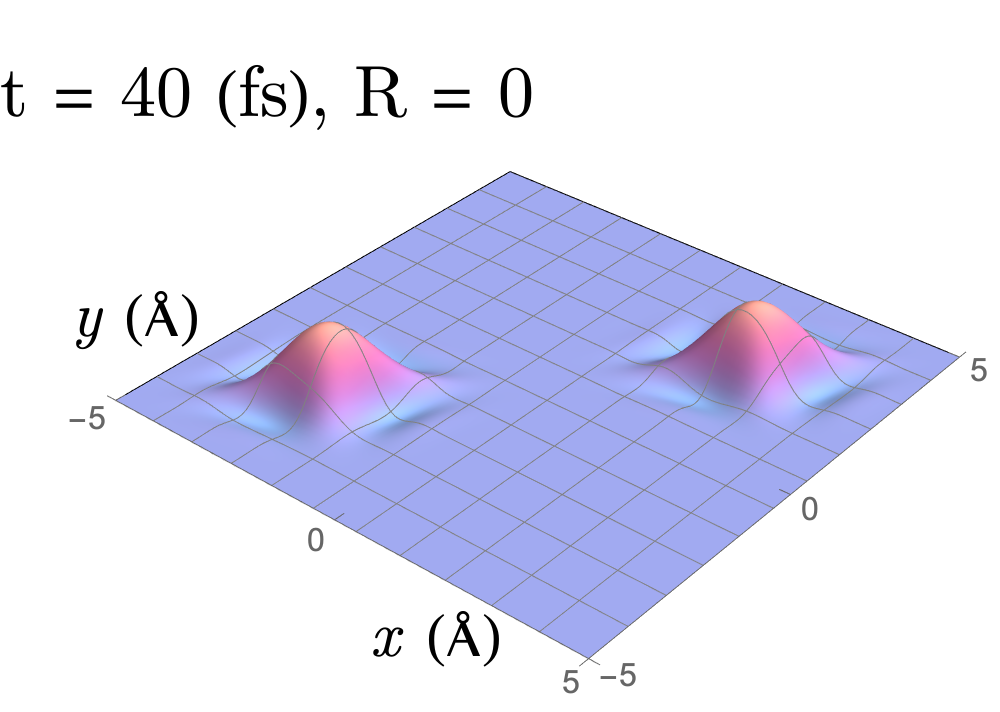} 
  \includegraphics[height=2.9cm]{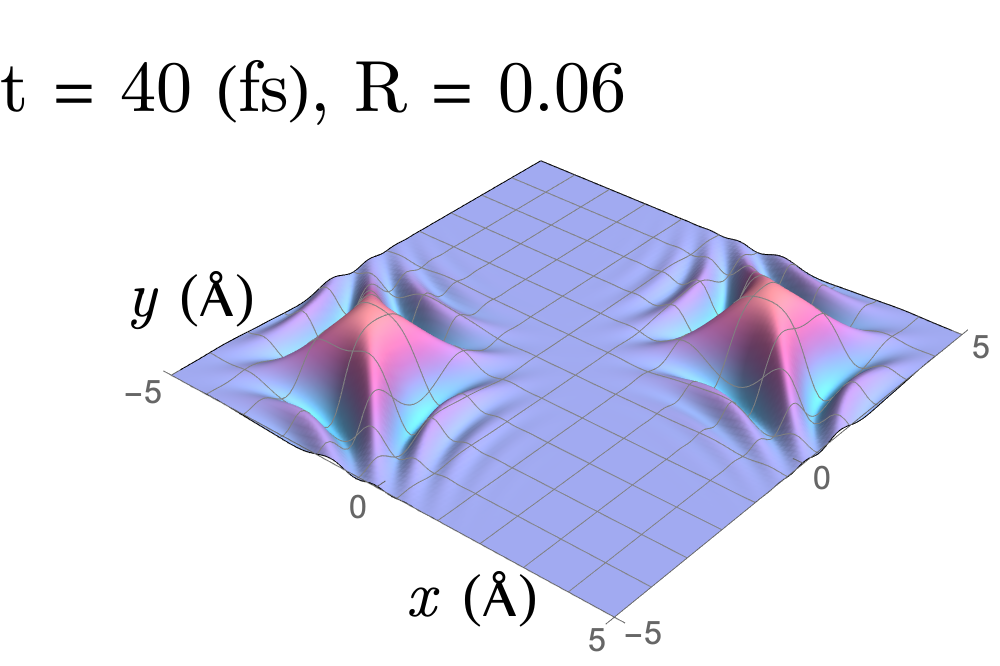}\\
  \includegraphics[height=2.9cm]{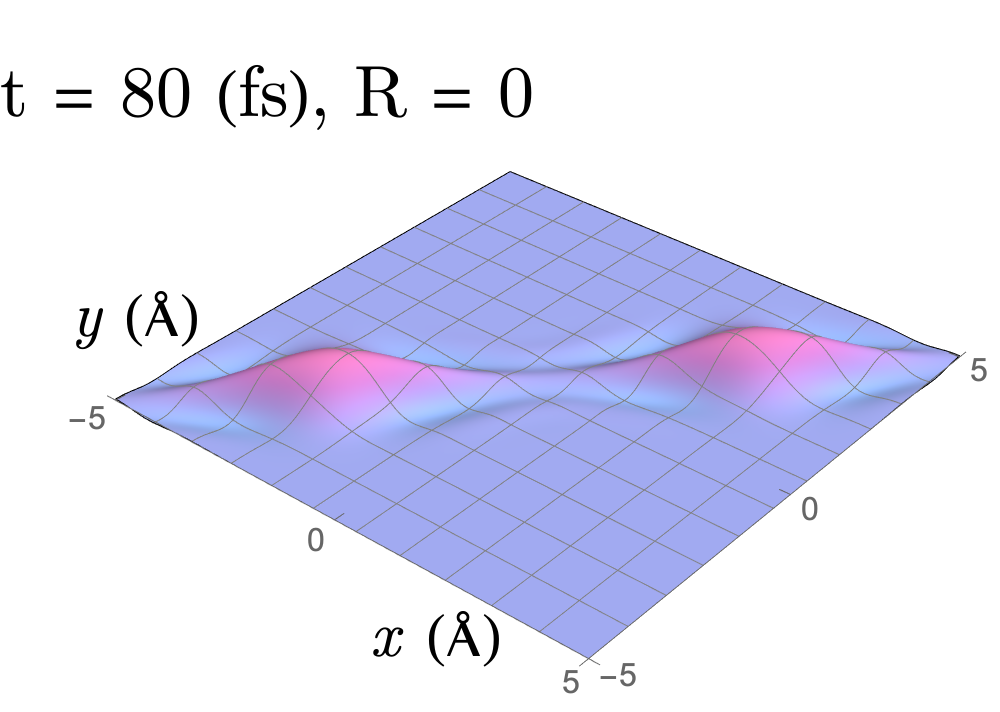} 
  \includegraphics[height=2.9cm]{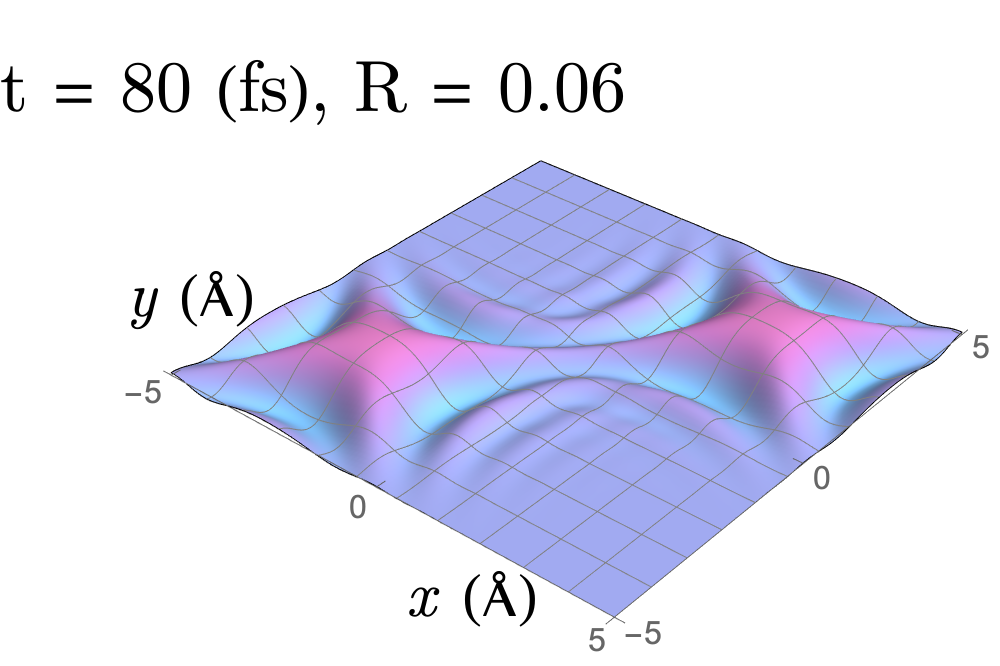}\\
  \includegraphics[height=2.85cm]{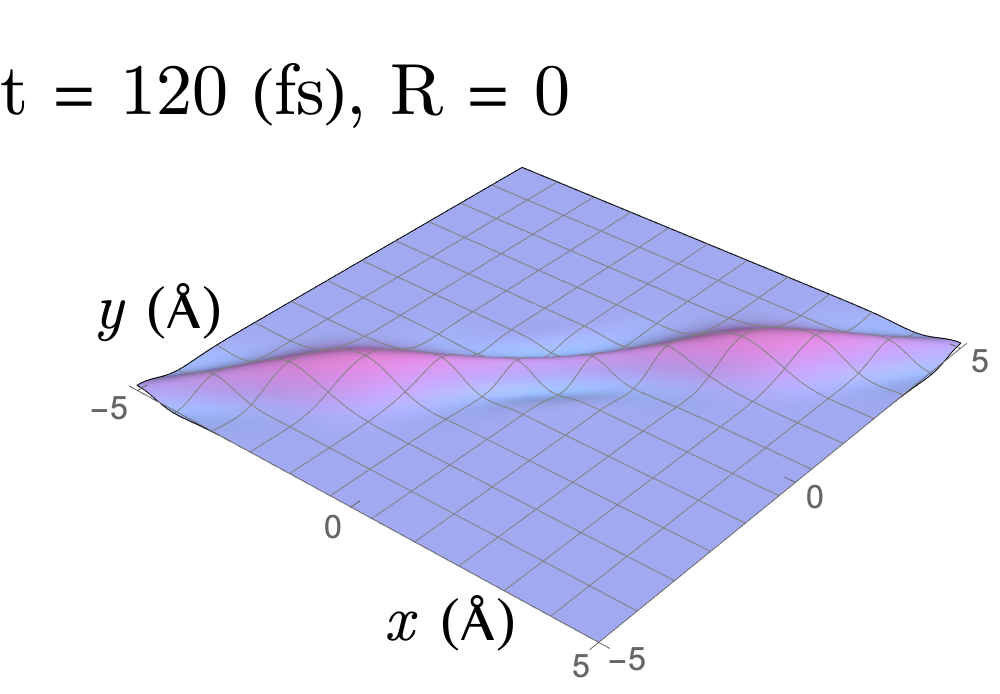} 
  \includegraphics[height=2.85cm]{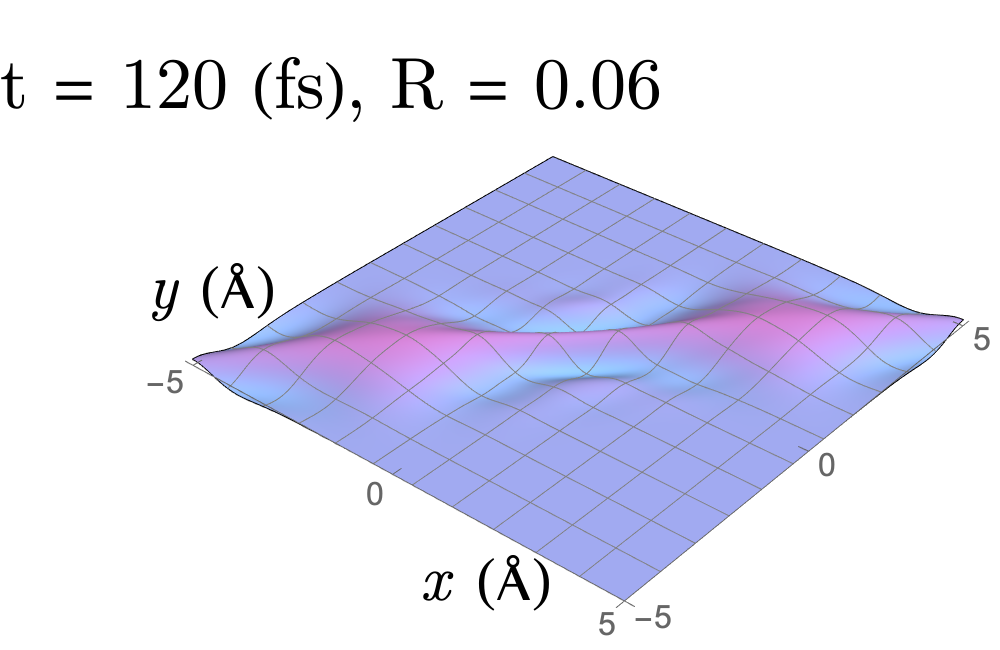}
  \caption{Medium time evolution of the density matrix, via Eq. in the Markovian case, $R_{\Omega}=0$ (left column), and the non-Markovian case, $R_{\Omega}=0.06$ (right column), for the free particle. Plots are over 120fs, to highlight the emergence of lateral coherences. Parameters are $t=320 {\rm K}$, $\gamma = \frac{k T}{20 \hbar}{\rm fs}^{-1}$. Significant deviation between the Markovian and non-Markovian cases is apparent in the emergence of lateral coherences, which spread the peaks of the density matrix orthogonal to the diagonal line.}\label{3DPlots_med}
  \end{figure}

We next analyse the non-Markovian behaviour on medium-to-long timescales, as shown in Fig. \ref{3DPlots_med}. This regime allows us to identify a more interesting role played by the Liouvillian  $\hat{L}_{\text{O}}$. Further to being responsible for a slower decoherence process at short times, it also causes the density matrix to spread away from the main diagonal after a longer time -- a process we will call {\em lateral coherences}. This phenomenon occurs long after the original off-diagonal coherences have decayed away and is a rather different signature of quantum behaviour than the coherence of the initial state.

\begin{figure} 
  \includegraphics[height=5.5 cm, angle=0]{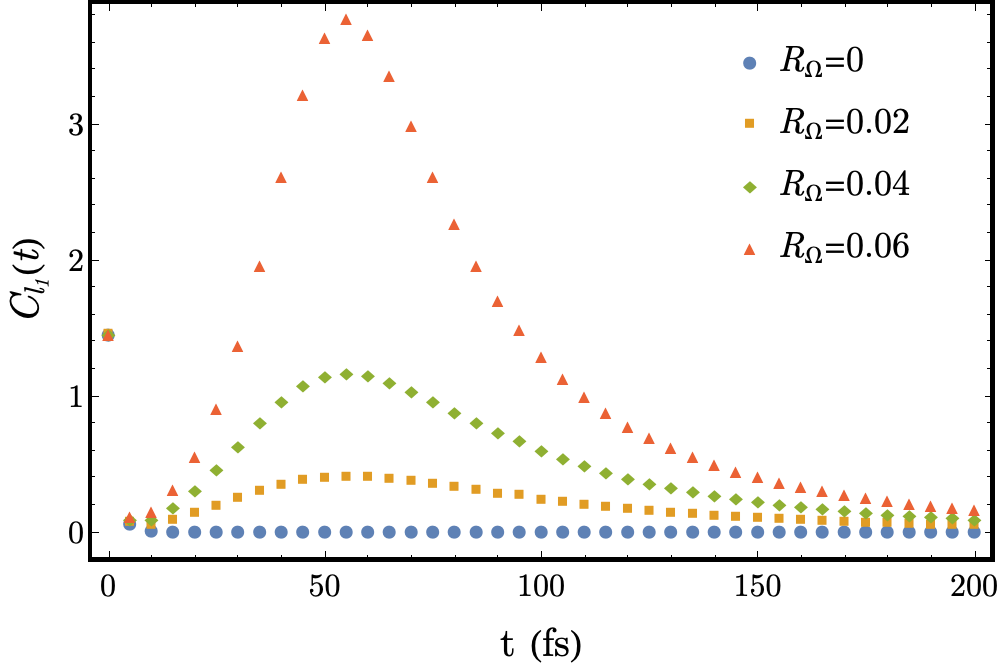}\\
  \caption{The time evolution of the $l_1$-norm of coherence, via Eq. (\ref{coheq}), over 200 fs. Parameters are $t=320 {\rm K}$, $\gamma = \frac{k T}{20 \hbar}{\rm fs}^{-1}$. The resurgence of coherence, more pronounced with increasing $R_{\Omega}$, is clearly observed. In the most extreme ($R_{\Omega}=0.06$) case, the coherence resurgence exceeds the initial value.}
  \label{decbehaviour}
  \end{figure}

Figure \ref{decbehaviour} shows the long time behaviour of the coherences in the system. A resurgence of coherent behaviour is evident, and can be associated with the emergence of lateral coherences identified in Figure \ref{3DPlots_med}. A variety of values of $R_{\Omega}$ are shown to demonstrate the varying magnitude of the coherence resurgence. In the most extreme case ($R_{\Omega} = 0.06$) the coherence resurgence overshoots the initial value.

We interpret the resurgence of these lateral coherences as a signature of the non-Markovian character of the system. Coherence can be linked to entropy in quantum systems -- for example via the relative entropy of coherence \cite{Baumgratz14}. The von Neumann entropy of a mixed Gaussian state is given in \cite{Horhammer08} as 
\be
\frac{S_{\rm vN}(\tilde{\rho})}{k_B} = \frac{1 - \mu}{2 \mu} \ln \frac{1 + \mu}{1 - \mu} - \ln \frac{2 \mu}{1 + \mu},
\ee
where $\mu = {\rm Tr}(\tilde{\rho}^2)$ is the purity. Therefore, the corresponding von Neumann entropy production rate is 
\be 
\frac{\dot{S}_{\rm vN}(\tilde{\rho})}{k_B} = - \frac{\dot{\mu}}{2 \mu^2} \ln \frac{1 + \mu}{1 - \mu}.
\ee

\begin{figure}[t] 
  \includegraphics[height=5.5 cm, angle=0]{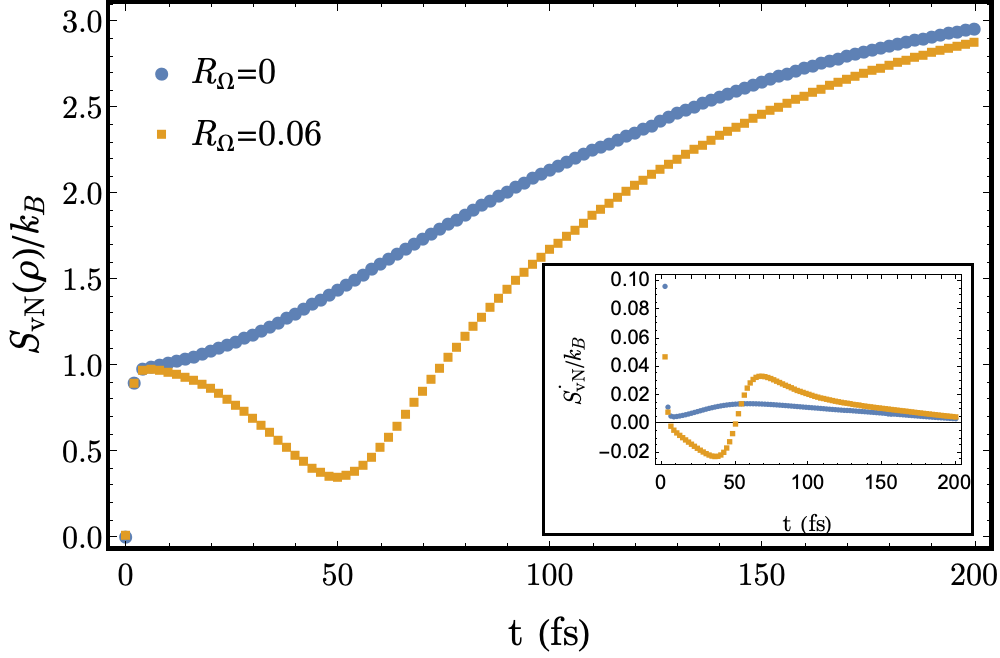}
  \caption{The von Neumann entropy of the density matrix, in units of $1 / k_B$, for the Markovian ($R=0$) and non-Markovian ($R=0.06$) cases. Parameters are $T=320 {\rm K}$, $\gamma = \frac{k T}{20 \hbar}{\rm fs}^{-1}$. In the non-Markovian case, there is a transient reduction of the system entropy, corresponding negative entropy production rate. Inset: entropy production rate. The non-Markovian entropy production rate clearly takes on negative values, whereas the Markovian entropy production is always positive.}\label{vNent}
  \end{figure}

By calculating the purity directly from the analytic solution (as further discussed in the next section), we demonstrate in Fig. \ref{vNent} that the coherence resurgence is associated with a transient negative entropy production rate. However, the second law is never violated as the global entropy production is always positive. It has been shown \cite{Thomas18, Debiossac20, Ford06} that non-Markovian dynamics can never be used to create thermal engines which violate the second law, as there is a thermodynamic cost associated with preparing states and removing the system from the bath. 

We interpret the negative entropy production rate in the light of a growing body of literature \cite{Strasberg19, Rivas20, Popovic18, Bhattacharya17, Marcantoni17} which characterises this kind of behaviour as characteristic of non-Markovianity. In the standard thermodynamic framework, entropy can be understood as measuring the amount of thermal energy within a system that is available to perform useful work. It has been shown that a so-called Szilard engine \cite{Kim11} can use quantum information to do work, e.g. via a Maxwell demon \cite{Cottet17}. By combining the thermodynamic and information theoretic viewpoints as in \cite{Ahmadi21}, which is necessary for a full understanding of quantum thermodynamics, we see that increasing entropy can be understood as a loss of information about the system's internal degrees of freedom. However, information backflow is considered characteristic of non-Markovianity \cite{Mazzola10, Bylicka16, Lu10, Chruscinski11}. When information flows back into the system, it can be interpreted via Szilard's framework as a new source of work. Therefore, there must be a corresponding entropy decrease to accommodate the system's increased thermodynamic capability. We further note that coherence itself can be a source for work \cite{Korzekwa16}, and therefore this interpretation can be understood in terms of either information or coherence.

\section{Tests of Positivity}

Density matrices must have spectra of positive eigenvalues summing to unity in order to consistently describe a quantum system. While it is well-known that master equations in the Lindblad form preserve the positivity of the density matrix \cite{Lindblad76}, the same cannot be said for the Caldeira-Leggett master equation \cite{Homa19}. Our novel master equation deviates even further from the Lindblad form because of its non-Markovian dynamics. As a consequence it is natural to ask whether or not positivity is preserved in our model.

In order to check whether the density matrix violates positivity as it evolves in time, we would ideally need to compute its eigenvalues, and verify that none of them becomes negative over time. Unfortunately this method is not straightforward when the density matrix represents a continuous system, as in our case, and so a different approach is taken. Thus we follow the approach presented in \cite{Homa19}, and compute the purity of the system and the Robertson-Schr\"odinger uncertainty relation. Strictly speaking, these quantities only provide necessary, but not sufficient, conditions for the positivity of the density matrix, and therefore must be taken only as an indication of whether any violation of positivity is occurring in the system. 

Let us consider first the purity of the density matrix, identified by the value ${\rm Tr}(\rho^2)$ where, for a pure state, ${\rm Tr}(\rho^2)=1$. For a mixed state, 
\be
\rho=\sum_i \lambda_i \vert \phi_i\rangle\langle \phi_i\vert\ ,
\ee
where $\lambda_i$ are the eigenvalues of $\rho$, with $\sum_i \lambda_i = 1$ and $0 \leq \lambda_i\leq 1$ for all $i$. Positivity of the density matrix is thus satisfied if 
\be
\sum_i\left(\lambda_i^2-\lambda_i\right) = {\rm Tr}(\rho^2) - {\rm Tr}(\rho) <0 .
\ee
Since ${\rm Tr}(\rho)=1$, positivity implies that ${\rm Tr}(\rho^2)<1$ at all times. Violation of this condition would imply violation of positivity of the density matrix, even though this condition being satisfied does not guarantee that the density matrix is positive definite \cite{Homa19}.  

\begin{figure}[t]
\includegraphics[width=6.5 cm]{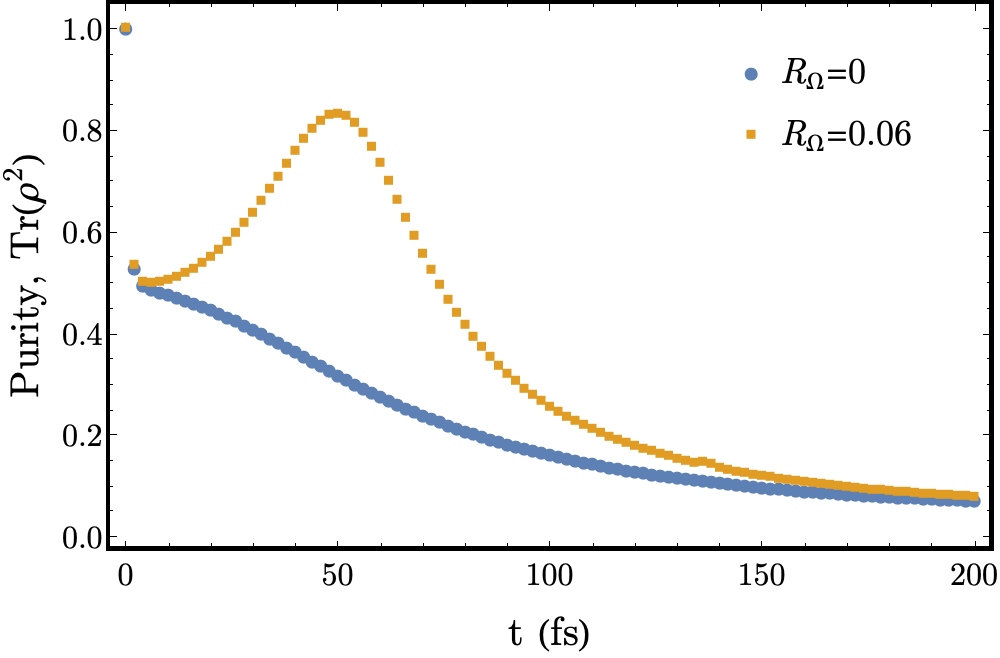}
\includegraphics[width=6.5 cm]{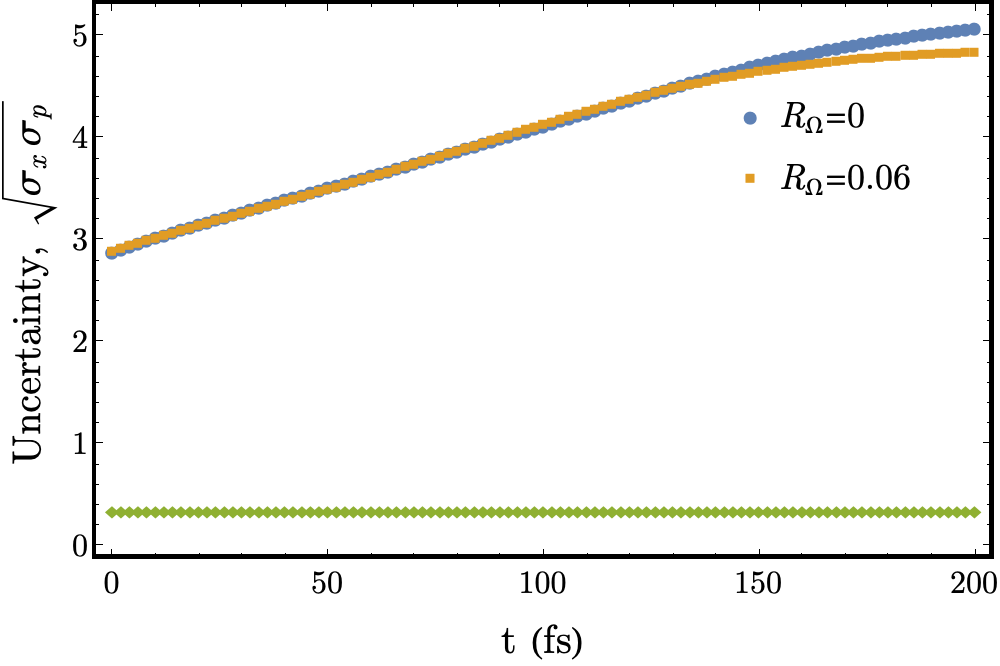}
\caption{Purity (top panel) and uncertainty (bottom panel) for Markovian $R_{\Omega}=0$ (blue) and non-Markovian $R_{\Omega} = 0.06$ (orange) cases for the free particle. Parameters are $T=320 {\rm K}$, $\gamma = \frac{k T}{20 \hbar}{\rm fs}^{-1}$.}
\label{a_01}
\end{figure}

Secondly, the Robertson-Schr\"odinger uncertainty relation reads
\be
\sigma_x \sigma_p \geq \frac{\hbar}{2}, \label{RS}
\ee
where
\be
  \sigma_x \sigma_p = \sqrt{\left({\rm Tr}(\rho \hat{x}^2)\!-\!{\rm Tr}^2(\rho \hat{x})\right)\!\left({\rm Tr}(\rho \hat{p}^2)\!-\!{\rm Tr}^2(\rho \hat{p})\right)},
 \label{Uceq}
 \ee
where $\hat{x}$ and $\hat{p}$ are the position and momentum operators. It has been shown \cite{Homa19} that the uncertainty relation (\ref{RS}) is based on the assumption of the density matrix being positive definite, and therefore violation of positivity may reflect in violation of Eq. (\ref{RS}).

The time evolution of the uncertainty, $\sigma_{x} \sigma_{p}$, and of the purity, ${\rm Tr}(\rho^2)$, are shown in Fig.\ref{a_01}. We observe that both the Markovian and non-Markovian systems satisfy these necessary positivity criteria. We note that the purity undergoes a non-monotonic decrease in the non-Markovian case, which corresponds to the coherence resurgence and transient negative entropy production. However, it never exceeds the initial value of $1$. The uncertainty shows little disparity between the two cases during the resurgence, but tends to a slightly different final value.

\section{Classical Limit}

The standard Markovian Caldeira and Leggett model can be shown to correspond in the classical limit to classical Brownian motion. By using the formalism of the Wigner transform, the Kramers Equation \cite{Gardiner85} can be derived, which corresponds to a Fokker-Planck equation in $(x,p)$ variables. Here, we derive the evolution equation for the Wigner function and the corresponding Fokker-Planck equation in the classical limit in the case of non-Markovian dynamics. 

We begin by Wigner transforming the non-Markovian quantum master equation (\ref{nonMME}). The Wigner function $W$ is defined as
\be
W(x,p,t)\! = \! \frac{1}{2\pi \hbar} \int_{-\infty}^{\infty} \!\!\! e^{-ipy/\hbar}\langle x + \frac{y}{2}|\tilde{\rho}|x - \frac{y}{2} \rangle dy 
\ee
and allows us to rewrite Eq.(\ref{nonMME}) as:
\be
\frac{\partial W}{\partial t} = \bigg[ \hat{L}_{\text{M}}^{(\text{W})} + \frac{\gamma}{\pi \Omega} \bigg(\hat{L}_{\text{R}}^{(\text{W})} +\hat{L}_{\text{V}}^{(\text{W})} + \hat{L}_{\text{O}}^{(\text{W})} \bigg) \bigg] W \ , \label{WnonMME}
\ee
where
\be
&&\hspace{-0.3cm}\hat{L}_{\text{M}}^{(\text{W})} W = -\frac{p}{M}\frac{\partial W}{\partial x} + \frac{dV}{dx} \frac{\partial W}{\partial p}\nonumber \\
&&\hspace{1.0cm} + 2 \gamma \frac{\partial (pW)}{\partial p} + 2 M \gamma k_{B} T \frac{\partial^{2} W}{\partial p^{2}} \ ,\\
&&\hspace{-0.3cm} \hat{L}_{\text{R}}^{(\text{W})} W =  4 \gamma \frac{\partial }{\partial p}(p W) \ , \label{WnonMME3}\\
&&\hspace{-0.3cm} \hat{L}_{\text{V}}^{(\text{W})} W =  - \sum^{\infty}_{n=o} \frac{\left( -\hbar^{2} \right)^{n}}{2^{2n}(2n)!}  \frac{d^{2n+1}V(x)}{d x^{2n+1}} \frac{\partial^{2n+1}W}{\partial p^{2n+1}} \ ,  \label{WnonMME2} \\
&&\hspace{-0.3cm} \hat{L}_{\text{O}}^{(\text{W})}W = 4 k_{B}T \frac{\partial^2 W}{\partial p \partial x} \ ,\label{WnonMME1} 
\ee
Notice that $\hat{L}_{\text{M}}^{(\text{W})}$ corresponds to the familiar Fokker-Planck operator describing classical Brownian motion.

By taking the classical limit, $\hbar \rightarrow 0$, we obtain an expression containing the standard Markovian Fokker-Planck equation, but with additional non-Markovian corrections:
\be
&&\hspace{-0.3cm}\frac{\partial W}{\partial t} = - \frac{p}{M}\frac{\partial W}{\partial x}\nonumber \\
&&\hspace{1.0cm} + (1+R_\Omega) \frac{dV}{dx}\frac{\partial W}{\partial p} + 2 \gamma (1+2 R_\Omega) \frac{\partial}{\partial p}(pW)\nonumber \\
&&\hspace{1.0cm}+ 2M\gamma k_BT \frac{\partial^2W}{\partial p^2}+ 4 k_BT R_\Omega \frac{\partial^2 W}{\partial x\partial p} . \label{FP}
\ee
Here the first four terms correspond to the familiar Markovian form, while the last term is purely non-Markovian. Notice that the relaxation and the potential terms are modified by corrective factors for finite $\Omega$, which allow us to reclaim the results of \cite{Caldeira83} in the Markov limit $\Omega \rightarrow \infty$.

The last, non-Markovian, term in the Fokker-Planck equation (\ref{FP}) describes novel dynamics brought about through non-Markovian memory effects. This term had already been identified in previous work \cite{Kohler97}, where it has been accounted for as part of the Markovian description of the damped harmonic oscillator.

It is interesting to note that our non-Markovian Fokker-Planck equation shares terms with the Markovian Fokker-Planck equation derived in \cite{Unruh89}, which describes the dissipative dynamics of a driven or damped harmonic oscillator. In particular, the term we label orthogonal which contains the mixed derivative $\partial_p \partial_q W$, and which is responsible for the novel non-Markovian dynamics, is also present in the dissipative damped quantum H-O Fokker-Planck equation. However, it is not an exact equivalence: the sign and magnitude of the term in our equation are both different, and our equation is not restricted to a choice of potential.

\section{Conclusions}

In this paper we have introduced a perturbative expansion method to derive quantum master equations describing non-Markovian Brownian Motion for any general potential acting on the system of interest. In contrast to \cite{Hu92,Fleming11b}, our derivation is not exact, as it is based on an expansion in $1/\Omega$, but it does provide a systematic way of addressing the time non-locality of the system. It is legitimate to expect that the expansion captures the relevant dynamics at large enough values of $\Omega$, while higher order terms can in principle be consistently computed according to needs.

We have also identified an exact analytical solution of the master equation here derived, whose numerical evaluation has provided us with interesting interpretation for the new term. While on short time scales, the standard decoherence process prevails, and the decay of the $l_1$-norm of coherences is not too different from the standard Markovian case, in the long time limit, dynamics seem to be much richer. A new phenomenon arises, which we name emergence of lateral coherences, by which a resurgence of coherent behaviour emerges orthogonally from the population structure of the density matrix. This coherence resurgence is associated with a transient negative entropy production rate, a phenomenon that is becoming recognised as characteristic of non-Markovian dynamics. In the resource theory framework \cite{Streltsov17}, coherence is a valuable tool to perform quantum operations. We speculate that the coherence resurgence may promote purely quantum behaviours, such as tunnelling, in non-Markovian open systems.

We have also discussed the positivity of the reduced density matrix under the evolution of our novel set of Liouvillians (\ref{nonMME1}--\ref{nonMME3}). While the argument that we adopt here is not strictly conclusive, it bears the same strengths and weaknesses presented in \cite{Homa19}. On the one hand analysing purity and uncertainty as proxies for positivity of the reduced density matrix is questionable in that the identified conditions for positivity violations are necessary but not sufficient. On the other hand, however, direct checks of positivity by explicit calculations of the eigenvalues of the density matrix are difficult, if not impossible, especially in the coordinate representation. Our findings show nonetheless that positivity violations are not detected in the parameter regime analysed here, despite being known to be present in general for this master equation \cite{Kohler97}. 

Finally, we have derived the corresponding Fokker-Planck equation in the classical limit. This clearly describes novel non-Markovian dynamics, relying on our conversion of non-local dynamics expressed in terms of convolution integrals into series of higher order spacial and momentum derivatives. While the methodology introduced here provides a clear advantage to effectively describe the non-Markovian dynamics involved in the system, it also has the drawback of deviating from more standard Langevin pictures for the corresponding classical system, leaving thereby questions open on the physical interpretation of the corresponding classical limit \cite{Kohler97,Cohen97}.

The non-Markovian framework introduced here is promising and worthy of further investigation. It potentially paves the way to possible applications in the broad area of quantum technologies, in all those cases in which harnessing quantum properties of the system of interest is desired, and to possible further understanding of the quantum-classical transition when memory effects are retained in the system of interest.\\

\noindent{\bf Acknowledgements}\\
Nick Werren acknowledges the University of Surrey for financial support. Sapphire Lally was supported by the Leverhulme Quantum Biology Doctoral Training Centre. 


\end{document}